\documentclass{Interspeech}

\interspeechcameraready

\title{Best Practices and Considerations for Child Speech Corpus Collection and Curation in Educational, Clinical, and Forensic Scenarios}

\author[affiliation={1}]{John}{H.L. Hansen$^*$}
\author[affiliation={1}]{Satwik}{Dutta$^*$}
\author[affiliation={2,3}]{Ellen}{Grand}

\affiliation{Center for Robust Speech Systems (CRSS)}{The University of Texas at Dallas}{USA}
\affiliation{}{National Crime Agency}{United Kingdom}
\affiliation{}{Lancaster University}{United Kingdom}

\email{john.hansen@utdallas.edu, satwik.dutta@utdallas.edu, e.grand@lancs.ac.uk}

\usepackage{comment}
\usepackage{setspace}
\begin{document}

\maketitle

\begin{abstract}
    
    A child’s spoken ability continues to change until their adult age. Until 7-8yrs, their speech sound development and language structure evolve rapidly. This dynamic shift in their spoken communication skills and data privacy make it challenging to curate technology-ready speech corpora for children. This study aims to bridge this gap and provide researchers and practitioners with the best practices and considerations for developing such a corpus based on an intended goal. Although primarily focused on educational goals, applications of child speech data have spread across fields including clinical and forensics fields. Motivated by this goal, we describe the WHO, WHAT, WHEN, and WHERE of data collection inspired by prior collection efforts and our experience/knowledge. We also provide a guide to establish collaboration, trust, and for navigating the human subjects research protocol. This study concludes with guidelines for corpus quality check, triage, and annotation.  
\end{abstract}

\section{Introduction}

\begingroup
\renewcommand{\thefootnote}{}% Remove number
\footnotetext{$^*$Equal Contribution. Supported by National Science Foundation.}
\endgroup

With the advent of data-driven technology, the availability of high-quality annotated data has become of paramount importance to AI/ML researchers. In the context of speech, the community has made consistent efforts to curate speech datasets over the last few decades. Efforts of multiple organizations such as LDC, NIST, MIT, and SRI culminated in the 1993 benchmark tests of the collected speech data quality and Automatic Speech Recognition (ASR) for the ARPA Spoken Language Program which included Wall Street Journal (WSJ) and Air Travel Information System (ATIS) datasets. In \cite{cieri2002research}, authors explained the challenges of conversational speech collection by three speech data types (broadcast news, telephone conversation, and meeting recordings) by ranking them by difficulty. The authors concluded meeting recordings to be the most difficult. Prior research on robustness and speaker variability in stress and noise also points to challenges when employing speech corpora that are not collected in consistent scenarios.  
Various factors play a role in determining how challenging speech data collection, curation, and annotation (CCA) can be. These include: environment, recording device/apparatus, movement, change of location, multimodality, informality, impromptu speech, overlapped speech, multiple speakers, and observer's/annotators paradox.   

Speech data CCA has evolved immensely since TIMIT in the 20th century to massive datasets (corpora) such as GigaSpeech 2, MLS, and People’s Speech. These massive datasets each contain around 30,000 hours (or more) of read/conversational speech for various speech processing tasks. In contrast, the MyST dataset \cite{ward2019my}, the known largest speech dataset in American English curated for children, contains around 230 hours of transcribed speech, 130x smaller than the People’s Speech dataset. This shows the disparity in the amount of transcribed speech data of children available to the community compared to that of adults. It is noted that the amount of data is not the only challenge. Other issues include additional protocols required to perform human subjects research with children, parental/appropriate consent, privacy issues, collection environment, ability of children to participate with a given set of collection guidelines (e.g., children are not as responsive to following adult corpus collection protocols, etc. It is even more challenging to setup and collect speech data of younger children. Apart from these, many of the environment and logistics factors noted earlier, that make speech data CCA challenging also hold true for children.

\vspace{-1.5pt}
The vision to develop child-centered speech technology \cite{potamianos1997automatic} is motivated by building a support system for parents, teachers, caregivers, and practitioners (speech language pathologist, child safety enforcement agencies, etc). Broadly the goal could be - (1)\textbf{Educational}: speech/language development, language learning, pronunciation improvement, special education, (2)\textbf{Clinical}: speech assessment, speech therapy, and (3)\textbf{Forensics}: resources to support child-based forensic voice interviews/analysis, child privacy protection in legal testimony. Most open-source child speech corpora \cite{ward2019my,shobaki2000ogi,batliner05b_interspeech,shankar2024jibo} are curated for ASR tasks, with audio collected via prompts or interactive systems to produce words/phrases. Although these include fewer younger children, CHILDES \cite{macwhinney2000childes} and TALKBANK \cite{MacWhinney2007} are prominent sources of longitudinal recordings (many with transcriptions) of young children and adults. Although these are not curated for speech processing tasks, \cite{kalanadhabhatta2024playlogue} benchmarked a few CHILDES datasets, and \cite{dutta24_odyssey} developed a joint speaker and language recognition model. 
% \cite{eskenazi2013crowdsourcing,berisha2024responsible,ng2024tutorial} - 
In this study, we gather knowledge from various child speech datasets and our experience in such data collection \cite{dutta22_interspeech,DUTTA2025103460} to document best practices and considerations that one needs to consider for effective child speech corpus CCA in educational, clinical, and forensic scenarios. The protocols, including additional resources, are publicly available on GitHub\footnote{https://github.com/SatwikDutta/All-in-one-Handbook-for-Child-Speech-Corpus}.
% The paper is structured as follows: \textcolor{red}{Satwik to add depending on space}.

\begin{table*}[th]
  \caption{Various (Supervised) Child Speech Corpus Details and accommodating Factors/Information by Age.}
  \label{tab:wwwh}
  \centering
  \begin{tabular}{lcccccl}
    \toprule
    \textbf{Corpus} & \textbf{Goal(s)} & \textbf{Language(s)} & \textbf{Cohort} & \textbf{Speech} & \textbf{Microphone} & \textbf{Stimuli/other}\\
     & &  & \textbf{Age(yr)} & \textbf{Type} & \textbf{Type} & \textbf{Considerations}\\
    \midrule
    Hoff \cite{tulloch2023filling} & $\xi$ & A.English & 2.5-3.5 & 2S & Wearable & Naturalistic toy/book sessions\\
    NITK Kids \cite{ramteke19_interspeech} & $\xi$ & Kannada & 2.5-6.5 & R;1S & Desk & Picture representing words\\
    Dutta et al. \cite{dutta22_interspeech}$^\cup$ & $\xi$,$\zeta$ & A.English & 3-5 & 2S & Wearable & Naturalistic adult-child childcare interactions\\
    kidsTALC \cite{rumberg22_interspeech} & $\xi$,$\zeta$ & German & 3-11 & 2S & Handheld & Questions on picture books by adult\\
    AusKidTalk \cite{ahmed21_interspeech} & $\xi$,$\zeta$ & Aus.English & 3-12 & R,1S & Headmount & Single words, sentences, narrative/emotional\\

    Cameron \cite{starke2023can} & $\xi$,$\zeta$ & A.English & 4-5 & 2S & Headmount & Book/play sessions, DAS-II, GFTA-3, Vocab test\\
    Jibo \cite{shankar2024jibo} & $\xi$ & A.English & 4-7 & R,1S & Robot & Letter, digit and oral discourse tasks\\
    ADOSMod3 \cite{lahiri23_interspeech} & $\zeta$ & A.English & 4-13 & 2S & - & ADOS, BOSCC for Autism\\
    PFSTAR \cite{batliner05b_interspeech} & $\xi$ & B. English,  & 4-14 & R,1S & Headmount & ITC-irst English Words and Phrases, AIBO, \\
    & & Ger.,Swe. & & & & Scribe sentences, Accent Diagnostic passage\\

    Kid Space \cite{aslan2024immersive} & $\xi$ & A.English & 6-7 & 2S & Wearable & Interactive multi-model learning platform\\
    Durante et al. \cite{durante2022causal} & $\int$ & - & 6-11 & 2S & - & Truth-telling and disclosure interviews\\
    Percept-R \cite{benway22_interspeech} & $\zeta$ & A.English & 6-24 & R & Headmount & Isolated syllables, words, multi-word phrases\\
    OGI \cite{shobaki2000ogi} & $\xi$ & A.English & 6-16 & R,1S & Headmount & Sti. displayed \& prompted, QA with human \\
    SingaKids \cite{chen16g_interspeech} & $\xi$,$\zeta$ & Man.Chinese & 7-12 & R & - & Phonetically balanced words, phrases, sentences\\
    MyST \cite{ward2019my} & $\xi$ & A.English & 8-11 & 1S & Headmount & Q\&A with virtual tutor\\
    \bottomrule
    \multicolumn{7}{l}{\textbf{Goal}: $\xi\rightarrow$Educational, $\zeta\rightarrow$Clinical, $\int\rightarrow$Forensics; \textbf{Type}:R=Read/Scripted,1S/2S=1-way/2-way Spontaneous;$^\cup\rightarrow$Uncontrolled.} \\
  \end{tabular}
\end{table*}

\section{Establishing the WHO, WHAT, HOW, and WHERE of the Data Collection}

Here, we discuss the WHO: child cohort selection, WHAT: type of speech data, HOW: the speech recording device/equipment and associated stimuli, and WHERE: the location/environment of speech data collection. The first question to ask before data collection is - ``\textbf{What is the research GOAL}?'' The goals of the research project will ultimately direct the data collection efforts. One primary factor for selection of the cohort is the age of the child followed by their ethnicity, race, gender, and languages spoken. Inspired by research in child speech development \cite{shriberg1993four}, we categorize speech sound development and language structure expected from children based on their age \textit{into four age groups}: \textbf{Early} (1-3), \textbf{Middle} (3-5), \textbf{Late} (5-8), and \textbf{Full Development} ($>$8).
% (see Table.\ref{tab:age_development}). 
It is also important to note whether child subjects have been diagnosed or are at-risk of speech, language or developmental delays. This would be crucial for speech assessment/therapy or other clinical tasks.  

% \begin{table}[th]
%   \caption{Comparison of speech sound development and language structure in children w.r.t age.}
%   \label{tab:age_development}
%   \centering
%   \begin{tabular}{ccc}
%     \toprule
%     \textbf{Age} & \textbf{Speech Sounds} & \textbf{Language}\\
%     \textbf{(yrs)}  & \textbf{Developing} & \textbf{Structure}\\
%     \midrule
%     1-3 & M “mama”, W “we”, HH “hi” & Vocalizations, \\
%     & B “baby”, D “daddy”, Y “you” & words \\
%     3-5 & T “two”, CH “chew”, JH “jump” & Words, \\
%     & K “cup”, F “fish”, V “van” &  sentences \\
%     5-8 & SH “sheep”, S “see”, Z “zoo”, & Full sentences \\
%     & R “red”, TH “think”, ZH “measure” & \\
%     $>$8 & Fully developed & Full sentences \\
%      & (typically developing child) & \\
%     \bottomrule
%   \end{tabular}
% \end{table}

Collected speech could be read/scripted or spontaneous 1-way or 2-way conversational interactions. Spontaneous speech could be child responses to open-ended questions or simply naturalistic conversational interactions. Responses to binary or polar questions can be considered as limited spontaneous speech. Spontaneous speech could also include dyadic conversations (between child and adult/robot/peer) and group-based interactions (preschool, classroom). Therefore, multi-speaker scenarios can include additional speakers (e.g. parents, teachers, caregivers, investigators, speech therapists, etc.). The voice capture/recording device is important. Traditionally for collecting clean speech, good quality head-mounted microphones connected to a device would be recommended. The device could be a smartphone, tablet/iPad, laptop, or desktop. Most modern devices have good quality in-built microphones, if the collection environment/room is noisefree, then noise canceling microphones might not be required. Omni-directional microphones are recommended for multi-speaker scenarios (e.g. positioned at opposite ends of a table) for the microphone to capture all speakers.  Head-mounted microphones would be connected to a desktop/laptop/tablet/iPad, smartphone, or wearable audio recording device. However, if children are young and collection environment is a classroom/childcare, wearable audio-recording devices such as LENA (\url{www.lena.org/}) are recommended. Other wearable options could include wireless clip-on microphones available from various vendors. Based on the project goals, investigators might also choose to collect multimodal data such as video of the recording environment or location and movement of subjects within the space. All video capturing devices have the ability to collect audio, where audio needs to be accessed by investigators. If investigators choose, they could record audio in parallel with multiple units in the environment, or positioned with multiple children. When various modalities of data are collected separately, it is important to time-mark to later align the data across modalities vs. time. Depending on scenario, recording location could be at home, research lab, childcare, or classroom. Studies considering child development, speech therapy, or clinical assessment could often consider a longitudinal based study as compared to a one time data collection. The standard time-gap for longitudinal studies depend on project goals, where collection needs may be staggered closely when children are younger (every 3 to 6 months) and could be widely spaced over time for older children (each year). Content stimuli would be required for read/scripted speech either in printed or on a screen for the subject. Various tools/prompts can be used to generate spontaneous speech such as STEM-based questions, interactive games, picture description tasks, play-based activities, etc. No stimuli or prompts are required for naturalistic conversations. 

For most research projects, meta-data will include a unique anonymized identifier for each subject followed by required recording/data details. Usually, this identifier helps identify the child without using their name (i.e. personal identifying info). Demographic meta-data would be: age, ethnicity, race, gender, languages spoken, known speech/other developmental delays, socio-economic status, and parents/guardian educational background. Experimental meta-data would include details about recording device, stimuli, and location as relevant to the study. It is crucial for investigators to know whether any particular challenges/issues emerge during a recording session to help avoid any discrepancies later. Forensic meta-data would consider and incorporate each of the aforementioned factors. Further to this, demographic information is vital for victim identification and safeguarding purposes. Aligning with the consideration for developmental delays, information collected on health status from a physical and mental perspective allows for appropriate measures and safeguards to be put in place in a forensic scenario and can assist the investigation with detailing factors which could be vital to the outcome. Attaining data from as broad a demographic group as possible ensures that the selected cross-section of participants are representative of society; this is a factor which is difficult to ensure in corpora of this nature due to the collection processes and time/cost constraints. 
% Previous corpora of child speech data (i.e. OGI, KidLUCID, CSLU Kids Corpus, MyST) often represent a more specific, and therefore less representative, demographic due to collections obtained from a single geographic location. In addition, recording device information such as; choice of microphone to create the best possible quality recording from which any degraded copies can be made to match casework, proximity between speaker and microphone. In addition to these recording factors such as; environmental (soft furnishings, empty room, outside, etc); ensuring EXIF data and file structure information is available for examination may also prove useful in the forensic analysis of audio data. Furthermore, being able to detect where audio may have been tampered with, whether spliced or otherwise altered, may be useful for forensic analyses of such data. 
When considering the type of conversation had, spontaneous and natural speech would be preferable as this would be the most representative of speaking styles encountered in audio examined for forensic purposes despite the challenges posed by this style of speaking. There is a significant lack of corpora, not only of children’s voices for younger ages, but that contains speech which is relevant for forensic examination. It is important to emphasize that if spontaneous speech can be produced then this can facilitate more forensically relevant research.

Data collection can either be: (i)\textit{supervised or unsupervised}: collection performed under a trained professional from the research team would be supervised, and (ii)\textit{controlled or uncontrolled}: collection in a sound booth or noise-free laboratory or room would be controlled, whereas collection in a classroom would be uncontrolled where the research team has limited control of the environment. While high-quality samples can be collected under controlled conditions, creating speech databases this way is challenging and impractical in most cases. Although not mandatory, young children will need additional supervision (such as parent/teacher/caregiver) and support during data collection efforts. Below we compare relevant details/info on various child speech corpora in Table \ref{tab:wwwh}.

\section{Pre-collection Steps}

\subsection{Navigating the Human Subjects Research Protocol}

Setting up human subjects research protocol through your Institutional Review Board (IRB) is mandatory for child speech data collection. The steps to establish the protocol can be challenging, and can take months before IRB approval. Investigators have ultimate responsibility for the study conduct, maintaining ethical standards for the study, and the protection of the rights and welfare of the subjects (children/families in this case) who participate in the study. They need to communicate clearly the study purpose, data collection requirements, where and how the data will be collected/stored/used, and what safeguards are in place to protect child privacy \cite{dutta24_syndata4genai}. \textit{A guide is available on GitHub}.

\subsection{Building Trust, Network and Relationship}

Spoken child data collection is not as simple as acquiring adult speech data. Additional efforts to obtain subject (child/parental) consent are mandatory. Subject recruitment is a challenge for any study involving children, however, many investigators do collaborate or seek partnerships with groups/teams that involve child activity. Investigators can also choose to recruit subjects independently without collaboration, however partnerships help streamline recruitment and help to ``get the word out''. Partnerships could include:(1)\textbf{Educational}: Teachers, Administrators, School Districts, Science Museums, Libraries, Tutoring Agencies/Volunteer Organizations, Researchers in Child Education, STEM Education,(2)\textbf{Clinical}: Clinical Research Centers, Hospitals, Researchers - Speech, Language and Hearing Sciences, Child Development, Early Childhood Intervention, and (3)\textbf{Forensics}: Local, state and federal Enforcement Agencies, and internationally between law enforcement/government agencies where collaboration is feasible. 

\section{Post-collection Steps}

\subsection{Triage \& Quality Check}

For data collection in supervised and controlled settings, a trained professional (from the investigator’s team) would verify recording equipment operations, check recordings are complete, and repeat if requirements are not satisfied. For data collection in supervised and uncontrolled settings, the trained professional would continue to perform necessary data collection operation checks, however environmental factors will add additional uncertainty in the recordings.  Collection in active classroom settings introduces uncontrolled acoustic events/noise in recorded data, wearable recorders might fail  or switch off due to unknown circumstances, etc. Investigators should be prepared to address such challenges. For Unsupervised Settings, the parent/teacher/caregiver (for young children) or minors (above 13 years) would need to be trained to ensure recordings meet the study guidelines.  Based on our prior experience, unsupervised data collection with young children is challenging because the adult (e.g., parent/teacher/caregiver) would need to both manage the learning/child environment as well as ensure reliable audio recordings. This goes both for controlled and uncontrolled settings. In this case, quality checks by the investigator/team should be to performed regularly or remotely at regular intervals across data collection cycle. We suggest audio recordings to be sampled at a minimum of 16 kHz, and encoded in 16-bit pulse-code modulation (PCM) format to be ideal. Development of modern speech technology applications rely on (large amount of) high-quality training data. A preliminary assessment of the collected speech data must be performed to make sure that the data is reliable and can be used. We suggest two tools for speech quality check: NIST STNR (Signal-To-Noise-Ratio) and Voice Activity Detection (VAD) based on prior research for data analysis \cite{mak2014study,ziaei2014speech,kumar2020improving}. NIST STNR and WADA STNR \cite{kim08e_interspeech} (better if $>$20dB) is a non-reference quality measure and higher values represent a better quality measurement. 
% Finally, while audio capture systems/recorders generally have an automatic gain control (AGC), they are not perfect and background acoustic events can cause major acoustic spikes which introduce unwanted clipping into the captured audio streams. In situations such as this, employing tools such as ClipDAT \cite{hansen2021nonlinear} can help assess the degree of nonlinear waveform clipping/distortion present in the audio streams.

\subsection{Human-in-the-Loop Data Annotation}

The three primary elements in traditional speaker-aware speech transcription are to: (i)identify the speaker, (ii)represent the spoken content,  and (iii)mark the spoken time. Transcription of read speech is the easiest,  where the only required task is to validate speech content with provided stimuli. Interactions with robots or virtual tutors are often recorded as individual utterances. However dyadic or multi-speaker naturalistic conversations with humans are generally long audio recordings requiring transcribers to individually map– speaker, text, and time.  This transcription challenge is not only elevated for spontaneous speech/multi-speaker scenarios, but is significantly more challenging for speech from young children in noisy environments. 

Many recent adult speech datasets rely on commercial speech transcription or data labeling platforms to estimate text content on a best-effort basis (online speech-to-text services). However, there can be prohibitions on sharing/uploading collected child speech data to such cloud platforms, since it is not always clear if they retain copies of the audio data. Careful attention to your institution’s IRB guidelines is critical (e.g., who has access to the data, where is human-subject data stored, etc.).  It is possible to use third-party cloud storage sites to help in sharing. Recruiting and training human transcribers is often a requirement for such tasks. For annotating spontaneous child speech data, it is recommended that annotators/transcribers have expertise in phonetics/linguistics. Although prior experience in data annotation and child speech development is not required, it is recommended based on our experience. One of the recommended tools for manual speaker-aware speech transcription would be the opens-source LDC Transcriber Tool1 \cite{barras1998transcriber} or it’s latest version TranscriberAG2. 

Discrete speaker tags, without using subject’s names, are usually assigned such as CHILD * and ADULT * where * denotes a number/key. In scenarios where the recording device is wearable, it is customary to annotate the subject who is wearing the device as the PRIMARY, FOCAL CHILD, or ADULT. All other speakers (in the vicinity) whose speech is captured by that recording device could be annotated as SECONDARY, NON-FOCAL CHILD, or ADULT. Considering various anomalies with naturalistic speech, there are best practices transcribers should consider for child speech.   Audio streams with long periods of silence,  or in uncontrolled naturalistic scenarios at home or classrooms could include environmental/background noise: appliances (fan, refrigerator), pets, media/music, furniture (moving chair, table), inaudible crowd noise, etc. Such non-speech sections should  be marked, and set aside  since they provide little information to the speech processing pipeline. Any personal identifiable information, if uttered, needs to be time-marked. This could include: names of children/adults/school/school district, components of address – street/county, etc. These portions of audio should be discarded for privacy and not used in the speech processing pipeline.  

A standard way to mark various anomalies or non-linguistic markers in naturalistic speech for transcribers is enclosing the event using a special code within brackets (e.g., “$[*]$” or “$<*>$”).  Annotators needs to mark words pronounced incorrectly by the child and truncated or fragmented words (e.g., “wh” for what, “pee” for pizza). Interjections, used to express a spontaneous feeling or reaction, are common in naturalistic speech especially of young children and needs to be marked given its importance in early childhood. This includes exclamation (ouch, wow), greetings (hey, bye), response (okay, oh, m-hm, huh), and hesitation (uh, er, um). Any breath related sounds such as cough, gasp, sneeze, etc. could be categorized with a single marker as they are not relevant and could be discarded for speech processing tasks.  When annotating children’s speech in noisy environments, annotation might face challenges comprehending the word spoken – we suggest that the annotators document their best guess in three tries and mark it. For any unintelligible or unknown word, annotators should mark them as ``OOV'' - out-of-vocabulary.  

Annotation becomes even more challenging with infants (birth to 18months) and toddlers (18months to 3yrs), which needs additional knowledge and  training.  For speech/vocalizations from infants and toddlers, traditional annotation rules generally will not apply and available speech technology pipeline needs to be reformulated and customized. One approach is  to use speech technology to automatically track Individual Growth and Development Indicators (IGDI) for infants and toddlers \cite{greenwood2022remote}, which are research-based assessment measures used throughout the U.S. in Early Head Start and community-based infant-toddler programs, and also in Australia and Portugal. These measures of early language \& literacy development include Early Communication Indicator (ECI) \cite{greenwood2013advancing} through their gestures, vocalizations, words, and sentences. The IGDI-ECI could also be used for non-English languages, and research has shown its efficacy for children with Autism Spectrum Syndrome \cite{nowell2024adapting},  with dual-language learners from Latinx backgrounds \cite{king2022exploring}, and cross-cultural exploration \cite{buzhardt2019cross}. Out of the four elements of the IGDI-ECI, three are speech-based. The first IDGI-ECI element is a “vocalization utterance” defined as a non-word or unintelligible verbal utterance voiced by the child. This includes laugh, fillers (“mm”, “huh”), various sounds such as animals (“moo” imitating a cow), transportation (“vroom” imitating a tractor), or others (“ah”, “da”, “eee”). It is important to note this does not include crying, involuntary noises (e.g., hiccups, burps, cough).  This is followed by the other two speech-based IGDI-ECI elements - “single-word utterance” and “multiple-word utterance”. The actual IDGI-ECI scoring does not depend of the content of the utterance, but the rate of each element per minute. Given the complexity of producing words, each single-word utterance is given a weight of two and multiple-word utterance is given a weight of three. Speech technology could complement this manual and intensive scoring strategy, which is often considered burdensome by early childhood researchers.  Overall, the IDGI-ECIs demonstrate both educational and clinical use especially in making individual intervention decisions. Although crying is discarded as an IDGI-ECI, it is a crucial indicator that offers valuable insights into the physical and mental conditions of infants, such as hunger, discomfort and pain. Research in audio-based infant/baby cry detection and classification \cite{yao2022infant,xia2024multi} has emerged. Another valuable measure in early childhood communication skills is “conversational turns” in dyadic conversations, which could be measured using speaker diarization technology. Conversational turns supports early cognitive development, enhances reading and vocabulary skills, and social-emotional development. LENA technology \cite{gilkerson2017mapping}, which is able to identify and differentiate adult speech, child speech, and tv/electronic noise, has validated measure of conversational turns and is broadly accepted in eight languages. 

Finally, to obtain an estimate for inter-transcriber reliability, researchers will often have multiple transcribers working on the transcription effort to transcribe a small amount of the same audio files, without the transcribers knowing they are either being assessed, or identifying which files are used. This allows for some degree of inter-labeler reliability. For adult read/prompted speech, inter-labeler reliability is generally around 7-10\%, while for spontaneous speech, inter-labeler reliability increases to 25\% or more. If investigators wish to open-source or share collected data with the research community, which is highly encouraged, they need approval from the IRB (Researchers could require outside researchers to satisfy a no-cost license agreement, which specifies no sharing of data by the licensee, which can help in privacy protection for child speech). However, there can be specific project challenges related to children, which need an independent external third-party review of the data to ensure that privacy of subjects is protected.   

\section{Conclusion}
This study aims to serve as a guide for investigators to navigate the collection, curation, and annotation of a child speech corpus. It outlines the various factors from cohort selection, recording device/environment, and stimuli, to human subjects protocol, triage of collected audio, and annotation. It also highlights the need to build trust, form a network, and sustain community relationships, especially for longitudinal data collection. For additional resources, please refer to our GitHub page. 

% \section{Acknowledgements}
% This work is supported by NSF Grants 1918032, 2234916, and 2341384.

\bibliographystyle{IEEEtran}
\bibliography{mybib}

\begin{thebibliography}{10}
\providecommand{\url}[1]{#1}
\csname url@samestyle\endcsname
\providecommand{\newblock}{\relax}
\providecommand{\bibinfo}[2]{#2}
\providecommand{\BIBentrySTDinterwordspacing}{\spaceskip=0pt\relax}
\providecommand{\BIBentryALTinterwordstretchfactor}{4}
\providecommand{\BIBentryALTinterwordspacing}{\spaceskip=\fontdimen2\font plus
\BIBentryALTinterwordstretchfactor\fontdimen3\font minus \fontdimen4\font\relax}
\providecommand{\BIBforeignlanguage}[2]{{%
\expandafter\ifx\csname l@#1\endcsname\relax
\typeout{** WARNING: IEEEtran.bst: No hyphenation pattern has been}%
\typeout{** loaded for the language `#1'. Using the pattern for}%
\typeout{** the default language instead.}%
\else
\language=\csname l@#1\endcsname
\fi
#2}}
\providecommand{\BIBdecl}{\relax}
\BIBdecl

\bibitem{cieri2002research}
C.~Cieri, D.~Miller, and K.~Walker, ``Research methodologies, observations and outcomes in (conversational) speech data collection,'' in \emph{Proc. HLT 2002}, 2002.

\bibitem{ward2019my}
W.~Ward, R.~Cole, and S.~Pradhan, ``My science tutor and the myst corpus,'' \emph{Boulder Learning Inc}, 2019.

\bibitem{potamianos1997automatic}
A.~Potamianos, S.~S. Narayanan, and S.~Lee, ``Automatic speech recognition for children.'' in \emph{Eurospeech}, vol.~97, 1997, pp. 2371--2374.

\bibitem{shobaki2000ogi}
K.~Shobaki, J.-P. Hosom, and R.~Cole, ``The ogi kids’ speech corpus and recognizers,'' in \emph{Proc. of ICSLP}, 2000, pp. 564--567.

\bibitem{batliner05b_interspeech}
A.~Batliner, M.~Blomberg, S.~D'Arcy, D.~Elenius, D.~Giuliani, M.~Gerosa, C.~Hacker, M.~Russell, S.~Steidl, and M.~Wong, ``The pf\_star children's speech corpus,'' in \emph{Interspeech 2005}, 2005, pp. 2761--2764.

\bibitem{shankar2024jibo}
N.~B. Shankar, A.~Afshan, A.~Johnson, A.~Mahapatra, A.~Martin, H.~Ni, H.~W. Park, M.~Q. Perez, G.~Yeung, A.~Bailey \emph{et~al.}, ``The jibo kids corpus: A speech dataset of child-robot interactions in a classroom environment,'' \emph{JASA Express Letters}, vol.~4, no.~11, 2024.

\bibitem{macwhinney2000childes}
B.~MacWhinney, \emph{The CHILDES project: The database}.\hskip 1em plus 0.5em minus 0.4em\relax Psychology Press, 2000, vol.~2.

\bibitem{MacWhinney2007}
------, \emph{The Talkbank Project}.\hskip 1em plus 0.5em minus 0.4em\relax London: Palgrave Macmillan UK, 2007, pp. 163--180.

\bibitem{kalanadhabhatta2024playlogue}
M.~Kalanadhabhatta, M.~M. Rastikerdar, T.~Rahman, A.~S. Grabell, and D.~Ganesan, ``Playlogue: Dataset and benchmarks for analyzing adult-child conversations during play,'' \emph{Proceedings of the ACM on Interactive, Mobile, Wearable and Ubiquitous Technologies}, vol.~8, no.~4, pp. 1--34, 2024.

\bibitem{dutta24_odyssey}
S.~Dutta, I.~López-Espejo, D.~Irvin, and J.~H.~L. Hansen, ``Joint language and speaker classification in naturalistic bilingual adult-toddler interactions,'' in \emph{The Speaker and Language Recognition Workshop (Odyssey 2024)}, 2024, pp. 81--85.

\bibitem{dutta22_interspeech}
S.~Dutta, S.~A. Tao, J.~C. Reyna, R.~E. Hacker, D.~W. Irvin, J.~F. Buzhardt, and J.~H. Hansen, ``Challenges remain in building asr for spontaneous preschool children speech in naturalistic educational environments,'' in \emph{Interspeech 2022}, 2022, pp. 4322--4326.

\bibitem{DUTTA2025103460}
S.~Dutta, D.~Irvin, and J.~H. Hansen, ``Exploring discrete speech units for privacy-preserving and efficient speech recognition for school-aged and preschool children,'' \emph{International Journal of Human-Computer Studies}, vol. 199, p. 103460, 2025.

\bibitem{tulloch2023filling}
M.~K. Tulloch and H.~Erika, ``Filling lexical gaps and more: code-switching for the power of expression by young bilinguals,'' \emph{Journal of Child Language}, vol.~50, no.~4, pp. 981--1004, 2023.

\bibitem{ramteke19_interspeech}
P.~B. Ramteke, S.~Supanekar, P.~Hegde, H.~Nelson, V.~Aithal, and S.~G. Koolagudi, ``Nitk kids’ speech corpus,'' in \emph{Interspeech 2019}, 2019, pp. 331--335.

\bibitem{rumberg22_interspeech}
L.~Rumberg, C.~Gebauer, H.~Ehlert, M.~Wallbaum, L.~Bornholt, J.~Ostermann, and U.~Lüdtke, ``kidstalc: A corpus of 3- to 11-year-old german children’s connected natural speech,'' in \emph{Interspeech 2022}, 2022, pp. 5160--5164.

\bibitem{ahmed21_interspeech}
B.~Ahmed and et~al., ``Auskidtalk: An auditory-visual corpus of 3- to 12-year-old australian children’s speech,'' in \emph{Interspeech 2021}, 2021, pp. 3680--3684.

\bibitem{starke2023can}
K.~Starke, ``“can i play with the toys now?”: An exploration of preschool storytelling structure in wordless picturebook and guided play contexts to inform teacher training and toy development,'' Ph.D. dissertation, State University of New York at Buffalo, 2023.

\bibitem{lahiri23_interspeech}
R.~Lahiri, T.~Feng, R.~Hebbar, C.~Lord, S.~H. Kim, and S.~Narayanan, ``Robust self supervised speech embeddings for child-adult classification in interactions involving children with autism,'' in \emph{Interspeech 2023}, 2023, pp. 3557--3561.

\bibitem{aslan2024immersive}
S.~Aslan, L.~M. Durham, N.~Alyuz, E.~Okur, S.~Sharma, C.~Savur, and L.~Nachman, ``Immersive multi-modal pedagogical conversational artificial intelligence for early childhood education: An exploratory case study in the wild,'' \emph{Computers and Education: Artificial Intelligence}, vol.~6, p. 100220, 2024.

\bibitem{durante2022causal}
Z.~Durante, V.~Ardulov, M.~Kumar, J.~Gongola, T.~Lyon, and S.~Narayanan, ``Causal indicators for assessing the truthfulness of child speech in forensic interviews,'' \emph{Computer speech \& language}, vol.~71, p. 101263, 2022.

\bibitem{benway22_interspeech}
N.~Benway, J.~L. Preston, E.~Hitchcock, A.~Salekin, H.~Sharma, and T.~McAllister, ``Percept-r: An open-access american english child/clinical speech corpus specialized for the audio classification of //,'' in \emph{Interspeech 2022}, 2022, pp. 3648--3652.

\bibitem{chen16g_interspeech}
N.~F. Chen, R.~Tong, D.~Wee, P.~Lee, B.~Ma, and H.~Li, ``Singakids-mandarin: Speech corpus of singaporean children speaking mandarin chinese,'' in \emph{Interspeech 2016}, 2016, pp. 1545--1549.

\bibitem{shriberg1993four}
L.~D. Shriberg, ``Four new speech and prosody-voice measures for genetics research and other studies in developmental phonological disorders,'' \emph{JSLHR}, vol.~36, no.~1, pp. 105--140, 1993.

\bibitem{dutta24_syndata4genai}
S.~Dutta and J.~H. Hansen, ``Navigating the united states legislative landscape on voice privacy: Existing laws, proposed bills, protection for children, and synthetic data for ai,'' in \emph{Synthetic Data’s Transformative Role in Foundational Speech Models}, 2024, pp. 91--95.

\bibitem{mak2014study}
M.-W. Mak and H.-B. Yu, ``A study of voice activity detection techniques for nist speaker recognition evaluations,'' \emph{Computer Speech \& Language}, vol.~28, no.~1, pp. 295--313, 2014.

\bibitem{ziaei2014speech}
A.~Ziaei, A.~Sangwan, and J.~H. Hansen, ``A speech system for estimating daily word counts,'' in \emph{Interspeech}, 2014.

\bibitem{kumar2020improving}
M.~Kumar, S.~H. Kim, C.~Lord, and S.~Narayanan, ``Improving speaker diarization for naturalistic child-adult conversational interactions using contextual information,'' \emph{The Journal of the Acoustical Society of America}, vol. 147, no.~2, pp. EL196--EL200, 2020.

\bibitem{kim08e_interspeech}
C.~Kim and R.~M. Stern, ``Robust signal-to-noise ratio estimation based on waveform amplitude distribution analysis,'' in \emph{Interspeech 2008}, 2008, pp. 2598--2601.

\bibitem{barras1998transcriber}
C.~Barras, E.~Geoffrois, Z.~Wu, and M.~Liberman, ``Transcriber: a free tool for segmenting, labeling and transcribing speech.'' in \emph{LREC}, vol.~98.\hskip 1em plus 0.5em minus 0.4em\relax Citeseer, 1998, pp. 28--30.

\bibitem{greenwood2022remote}
C.~R. Greenwood and et~al., ``Remote use of individual growth and development indicators (igdis) for infants and toddlers,'' \emph{Journal of Early Intervention}, vol.~44, no.~2, pp. 168--189, 2022.

\bibitem{greenwood2013advancing}
------, ``Advancing the construct validity of the early communication indicator (eci) for infants and toddlers: Equivalence of growth trajectories across two early head start samples,'' \emph{ECRQ}, vol.~28, no.~4, pp. 743--758, 2013.

\bibitem{nowell2024adapting}
S.~Nowell and et~al., ``Adapting the early communication indicator as a social communication outcome measure for young autistic children: A pilot study,'' \emph{AJSLP}, vol.~33, no.~5, pp. 2610--2617, 2024.

\bibitem{king2022exploring}
M.~King, A.~L. Larson, and J.~Buzhardt, ``Exploring the classification accuracy of the early communication indicator (eci) with dual-language learners from latinx backgrounds,'' \emph{Assessment for Effective Intervention}, vol.~47, no.~4, pp. 209--219, 2022.

\bibitem{buzhardt2019cross}
J.~Buzhardt and et~al., ``Cross-cultural exploration of growth in expressive communication of english-speaking infants and toddlers,'' \emph{ECRQ}, vol.~48, pp. 284--294, 2019.

\bibitem{yao2022infant}
X.~Yao, M.~Micheletti, M.~Johnson, E.~Thomaz, and K.~de~Barbaro, ``Infant crying detection in real-world environments,'' in \emph{ICASSP}.\hskip 1em plus 0.5em minus 0.4em\relax IEEE, 2022, pp. 131--135.

\bibitem{xia2024multi}
M.~Xia, D.~Huang, and W.~Wang, ``Multi-task learning for audio-based infant cry detection and reasoning,'' \emph{IEEE Journal of Biomedical and Health Informatics}, 2024.

\bibitem{gilkerson2017mapping}
J.~Gilkerson and et~al., ``Mapping the early language environment using all-day recordings and automated analysis,'' \emph{AJSLP}, vol.~26, no.~2, pp. 248--265, 2017.

\end{thebibliography}

\end{document}